\documentclass[aps,prd,showpacs,amsmath,10pt,twocolumn,superscriptaddress,floatfix,nofootinbib,notitlepage]{revtex4-2}
\usepackage[utf8]{inputenc}
\usepackage[english]{babel}
\usepackage[dvipsnames]{xcolor}
\usepackage[normalem]{ulem}
\usepackage[
colorlinks=true,
linkcolor=blue,
breaklinks=true,
urlcolor=magenta,
citecolor=blue]{hyperref}

\usepackage{bm,color,xcolor,amsfonts,amsmath,amssymb,epsfig,esint,orcidlink}

\newcommand{\Eth}{E^{\rm thr}}
\newcommand{\be}{\begin{equation}}
\newcommand{\ee}{\end{equation}}
\newcommand{\bea}{\begin{eqnarray}}
\newcommand{\eea}{\end{eqnarray}}
\newcommand{\beas}{\begin{eqnarray*}}
\newcommand{\eeas}{\end{eqnarray*}}

\newcommand{\Amp}{\mathcal{A}}

\newcommand{\rub}{\affiliation{Institut f\"ur Theoretische Physik II, Ruhr-Universit\"at Bochum, D-44780 Bochum, Germany }}

\newcommand{\fzj}{\affiliation{Institute for Advanced Simulation, Forschungszentrum J\"ulich, D-52425 J\"ulich, Germany}}

\newcommand{\itp}{\affiliation{CAS Key Laboratory of Theoretical Physics, Institute of Theoretical Physics, \\Chinese Academy of Sciences, Beijing 100190, China}}

\newcommand{\ucas}{\affiliation{School of Physical Sciences, University of Chinese Academy of Sciences, Beijing 100049, China}}

\newcommand{\JSI}{\affiliation{Jozef Stefan Institute, Jamova 39, 1000 Ljubljana, Slovenia}}

\newcommand{\lis}{\affiliation{CeFEMA, Center of Physics and Engineering of Advanced Materials, Instituto Superior T{\'e}cnico, Avenida Rovisco Pais 1, 1049-001 Lisboa, Portugal}}

\newcommand{\peng}{\affiliation{Peng Huanwu Collaborative Center for Research and Education, Beihang University, Beijing 100191, China}}

\newcommand{\scnt}{\affiliation{Southern Center for Nuclear-Science Theory (SCNT), Institute of Modern Physics,\\ Chinese Academy of Sciences, Huizhou 516000, China}}

\begin{document}

\title{How does the $X(3872)$ show up in $e^+e^-$ collisions: Dip versus peak}

\author{Vadim Baru\orcidlink{0000-0001-6472-1008}}\email{vadim.baru@tp2.rub.de}
\rub

\author{Feng-Kun~Guo\orcidlink{0000-0002-2919-2064}}\email{fkguo@itp.ac.cn}
\itp \ucas \peng \scnt

\author{Christoph~Hanhart\orcidlink{0000-0002-3509-2473}}\email{c.hanhart@fz-juelich.de}
\fzj

\author{Alexey Nefediev\orcidlink{0000-0002-9988-9430}}\email{Alexey.Nefediev@ijs.si}
\JSI
\lis

\begin{abstract}
We demonstrate that the dip observed near the total
energy of 3872 MeV in the recent cross section data from the BESIII Collaboration for $e^+e^-\to J/\psi\pi^+\pi^- $ admits a natural explanation as a coupled-channel effect: it is a consequence of unitarity and a strong $S$-wave $D\bar D^*$ attraction
that generates the state $X(3872)$. We anticipate the appearance of a similar dip in the $e^+e^-\to J/\psi\pi^+\pi^-\pi^0$ final state near the $D^*\bar{D}^*$ threshold driven by the same general mechanism, then to be interpreted as a signature of the predicted spin-2 partner of the $X(3872)$.
\end{abstract}

\maketitle

Since the first exotic state $\chi_{c1}(3872)$ [also known as $X(3872)$] in the spectrum of charmonium was discovered by the Belle Collaboration in 2003 \cite{Belle:2003nnu}, 
quarkoniumlike states have been under intensive studies as they provide a unique laboratory to test our understanding of excited hadrons and thus the nonperturbative regime of quantum chromodynamics (for reviews, we refer to Refs.~\cite{Hosaka:2016pey, Lebed:2016hpi, Esposito:2016noz, Guo:2017jvc, Olsen:2017bmm, Kalashnikova:2018vkv, Brambilla:2019esw, Chen:2022asf}).
Recent studies of the electron-positron annihilation in the energy range $3.8-3.9$~GeV performed by the BESIII Collaboration revealed the appearance of a dip rather than peak in the line shape in the vicinity of the mass of $X(3872)$~\cite{BESIII:2022ner}.\footnote{
The same data point is also seen as a dip at the energy 3871.3~MeV in the data in Ref.~\cite{BESIII:2022qal}. No signal around the $X(3872)$ mass was seen in the $J / \psi\pi^{+} \pi^{-}$ invariant mass distribution of the initial-state radiation process $e^{+} e^{-} \rightarrow \gamma_{\mathrm{ISR}} J / \psi\pi^{+} \pi^{-} $ due to the low statistics~\cite{BESIII:2015klj}.} We demonstrate in this Letter that this structure finds a natural explanation in the interplay of different production mechanisms operative for the $X(3872)$ production in $e^+e^-$ collisions as a concrete realisation of the universal mechanism introduced in Ref.~\cite{Dong:2020hxe}.

The scene is set by two competing production channels, for definiteness referred to as channel 1 and channel 2, with the thresholds $\Eth_1$ and $\Eth_2$, respectively. In what follows, 
$\Eth_2-\Eth_1>0$, and the energy $E$ is counted from the higher threshold, $\Eth_2$. 
We introduce 
\begin{itemize}
\item $a_{11}$ as a parameter that governs the single-channel interaction strength in channel 1 at the threshold of channel 2;
\item $a_{22}$ as the $S$-wave scattering length in channel 2 in case it is completely decoupled from channel 1;
\item $a_{12}$ as the parameter that describes coupling of channels 1 and 2.
\end{itemize} 
Then, the expression for 
the elastic scattering amplitude in channel 1 within the energy region around $\Eth_2$ 
obtained in a non-relativistic 
effective field theory for channel 2 reads~\cite{Dong:2020hxe}
\be
T_{11}(E)=
-8\pi \Eth_2\Bigg(\frac{1}{a_{11}^{-1}-ik_1}+\frac{a_{12}^{-2}(a_{11}^{-1}-ik_1)^{-2}}{a_{22, \mathrm{eff}}^{-1}-ik_2} \Bigg)
, \label{eq:t11}
\ee
where $k_1$ is the centre-of-mass momentum in channel 1, $k_2 \approx \sqrt{2\mu_2 E}$ is the momentum in channel 2 treated nonrealativisitcally in the energy range of interest,
$\mu_2$ is the reduced mass in channel 2, and the effective scattering length in channel 2 coupled to channel 1, $a_{22,\mathrm{eff}}$, is given by
\be
a_{22,\mathrm{eff}}^{-1}=a_{22}^{-1}-a_{12}^{-2}(a_{11}^{-1}-ik_1)^{-1}.
\label{eq:a22eff}
\ee
Notice that here channel 1 is not required to be nonrelativistic though the amplitude $T_{11}(E)$ takes the same form as that obtained when both channels are nonrelativistic~\cite{Cohen:2004kf}.

The first term on the right-hand side in Eq.~(\ref{eq:t11}) can be regarded as a background coming from channel 1 alone.
The channel coupling induces an interference, with the relative phase completely fixed by unitarity, that ensures the emergence of a dip in the line shape $|T_{11}|^2$ as long as $|a_{22}|$ is large enough. To see it explicitly, we employ relation (\ref{eq:a22eff}) to bring the amplitude in Eq.~(\ref{eq:t11}) to the form
\be
T_{11}(E)=\frac{-8 \pi \Eth_2\left(a_{22}^{-1}-i \sqrt{2 \mu_2 E}\right)}{\left(a_{11}^{-1}-i k_1\right)\left(a_{22, \mathrm{eff}}^{-1}-i \sqrt{2 \mu_2 E}\right)},
\label{eq:t11v2}
\ee
where the numerator on the right-hand side vanishes at $E=0$ for
$a_{22}\to\infty$,
the so-called unitary limit, 
while the denominator remains
finite in this limit.
The zero of $T_{11}$ in Eq.~\eqref{eq:t11v2} is an example of a Castillejo-Dalitz-Dyson zero~\cite{Castillejo:1955ed}; see also Ref.~\cite{Hanhart:2011jz}.
In a more realistic situation, when the interaction
in the decoupled channel 2 approaches the unitary limit (large $|a_{22}|$), the scattering amplitude in channel 1 must show a dip at the threshold of channel 2.

\begin{figure*}[t!]
\centering
\includegraphics[width=0.7\textwidth]{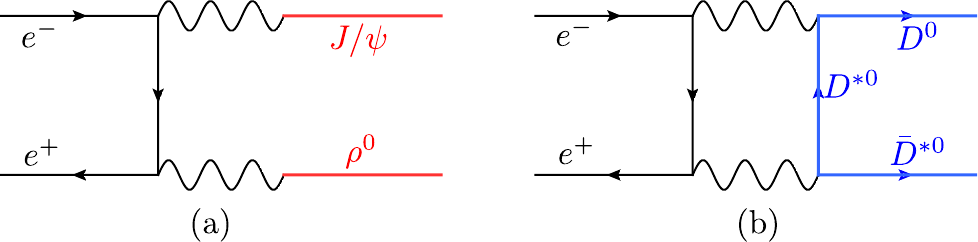}
\caption{The lowest-order diagrams for the $e^+e^-\to J/\psi\rho^0$ [diagram (a)] and $e^+e^-\to D^0\bar D^{*0}$ [diagram (b)] annihilation via two photons.}
\label{fig:eediagrams}
\end{figure*}

This universal picture allows one to interpret the well-known fact that the $f_0(980)$ appears as a dip in the $\pi\pi\to\pi\pi$ scattering amplitude (see, for example, Ref.~\cite{Dai:2014zta}), where $\pi\pi$ and $K\bar K$ act as channels 1 and 2, respectively, as a necessary consequence of the strong $S$-wave $K\bar K$ attraction.

Similarly, a dip at an energy around 1.67~GeV in the $K^{-} p \rightarrow K^{-} p$ and $K^{-} p \rightarrow \bar{K}^0 n$ total cross sections (see the data compiled in Ref.~\cite{Kamano:2015hxa})
 may indicate a strong $S$-wave attraction in the $\Lambda\eta$ channel, whose threshold is at 1664~MeV.
Further examples of near-threshold dip structures due to the same mechanism include a dip around the $\bar KN$ threshold in the $\pi\Sigma\to\pi\Sigma$ scattering amplitude from unitarised chiral perturbation theory (UChPT)~\cite{Oller:2000fj,Jido:2003cb} or lattice quantum chromodynamics~\cite{BaryonScatteringBaSc:2023zvt}, 
a dip around the $D_s^*\bar K$ threshold in the $D^*\pi\to D^*\pi$ scattering amplitude from UChPT~\cite{Guo:2006rp}, and so on.
Below we show that the same mechanism
is at work in the direct production of the $X(3872)$ in $e^+e^-$ collisions (see also Ref.~\cite{Denig:2014fha} for an estimate of the $X(3872)$ dielectron width). 

Recently the cross section of the reaction
$e^+e^-\to J/\psi\pi^+\pi^-$ was
measured by the BESIII Collaboration at several energies~\cite{BESIII:2022ner}, and
contrary to naive expectations, no enhancement around the $X(3872)$ mass was observed. Instead, there is an indication of a dip, 
although the data
are admittedly also consistent with
a flat distribution~\cite{BESIII:2022ner}. Below, we provide theoretical arguments supporting the necessity for the appearance of this dip and emphasise the importance of further
detailed studies to understand its manifestation in the data. 

We denote $J/\psi\rho$ and $D\bar D^*$ as channel 1 and channel 2, respectively, and neglect the $J/\psi\omega$ channel for simplicity. 
We note that the $J/\psi \rho^0$ state with $J^{PC}=1^{++}$ can be produced in $e^+e^-$ collisions at tree level through two virtual photons [see Fig.~\ref{fig:eediagrams}(a)] while the state $D\bar D^*+c.c.$ with the same quantum numbers can only be produced at the loop level [see Fig.~\ref{fig:eediagrams}(b)], thus the latter is expected to be suppressed by the geometric factor $1/(16\pi^2)$.\footnote{We also recall that the magnetic vertex $\gamma D^0\bar{D}^{*0}$ is proportional to the $D^0$ momentum that vanishes at the $D^0\bar{D}^{*0}$ (channel 2) threshold.} Therefore, we neglect the direct production through the $D\bar D^*+c.c.$ channel and write the $e^+e^-\to J/\psi\rho^0$ amplitude in the vicinity of the $D\bar D^*$ threshold as
\be
\Amp(\sqrt{s})=P_0+P_1T_{11}(E),
\label{eq:ee2MM}
\ee
where $\sqrt{s}=m_{D^0}+m_{D^{*0}}+E$ is the $e^+e^-$ centre-of-mass energy and the last term on the right-hand side describes rescatterings. In particular, it contains the $J/\psi \rho$-$D\bar D^*$ coupled-channel dynamics discussed above. Then the $J/\psi\rho\to J/\psi\rho$ scattering amplitude $T_{11}$ can be approximated by Eq.~\eqref{eq:t11v2} and the corresponding cross section reads
\be
\sigma(\sqrt{s})={\cal N}_0\int_{-\infty}^{+\infty} dw\, \frac{\left|\Amp(\sqrt{s}-w)\right|^2}{\sqrt{2 \pi}\delta_E} \exp \left(-\frac{w^2}{2 \delta_E^2}\right),
\label{eq:xsection}
\ee
where ${\cal N}_0$ is an overall normalisation factor and the signal is convolved with a Gaussian-distributed energy spread to mimic the
actual situation of the BESIII experiment, with the energy spread $\delta_E=1.7$~MeV~\cite{BESIII:2022ner}. Since the energy range of interest is narrow (from 3.8 to
3.9~GeV), only the leading energy dependence is retained in Eq.~(\ref{eq:xsection}) while the overall kinematical factor is treated as a constant for simplicity.

Since $P_0$ can be absorbed by the overall normalisation, ${\cal N}\equiv P_0^2{\cal N}_0$, the resulting model depends on five real parameters: $\{a_{11},a_{12},a_{22},R,{\cal N}\}$, where $R\equiv P_1/P_0$. Using the 
results of Ref.~\cite{Baru:2021ldu} and the data in Ref.~\cite{LHCb:2020xds}, the $D\bar D^*$ scattering length in the $X(3872)$ channel is determined to be
\be
a_{22,\text{eff}}=(-6.39+i11.74)~\mbox{fm} \, .
\label{eq:a22effN}
\ee
Then, in the considered two-channel formalism, Eq.~\eqref{eq:a22eff} constrains two real parameters (we choose them to be $a_{11}$ and $a_{12}$) through the third one ($a_{22}$). In particular, we use
\bea\label{Eq:a11}
a_{11}^{-1}=k_1\,\frac{{\rm Re}(a_{22,\text{eff}}^{-1})-a_{22}^{-1}}{{\rm Im}(a_{22,\text{eff}}^{-1})}
\eea
in the amplitude $T_{11}$ in Eq.~(\ref{eq:t11v2}) while the channel-coupling parameter can be obtained as
\begin{equation}\label{Eq:a12}
    a_{12}^{-1}=\sqrt{\frac{k_1}{{\rm Im}(a_{22,\text{eff}})}}\, \left|1-\frac{a_{22, \mathrm{eff}}}{a_{22}}\right|.
\end{equation}
To take into account a finite width of the $\rho$ meson, we evaluate the momentum $k_1$ as \cite{woolly,Braaten:2007dw}
\be 
k_1=\mbox{Re}\sqrt{\frac{[s{-}(m_{J/\psi}{+}m_\rho)^2][s{-}(m_{J/\psi}{-}m_\rho)^2]}{4s}} \, ,
\label{eq:k1}
\ee
with $m_\rho = (775-i75)$~MeV, and it is evaluated at the $D^0\bar D^{*0}$ threshold in Eqs.~\eqref{Eq:a11} and \eqref{Eq:a12}.
Notice that to fit to the BESIII data over the energy range from about 3.81 to 3.9~GeV, we keep the full expression for $k_1$ in Eq.~\eqref{eq:k1} and do not expand it in powers of $E$.

\begin{figure}[tb]
\centering
\includegraphics[width=\columnwidth]{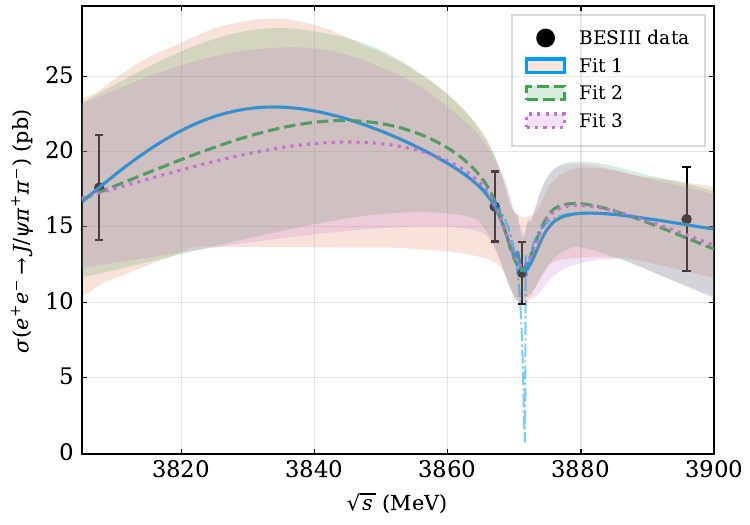}
\caption{The line shapes for the three best fits in Table~\ref{tab:fits} to the BESIII data~\cite{BESIII:2022ner} for the $e^+e^-\to J/\psi\pi^+\pi^-$ annihilation after convolution with the energy spread function---see Eq.~(\ref{eq:xsection}). As an example, the blue dashed line shows the line shape for fit~1 without the effect of the energy spread. The $1\sigma$ error bands correspond to the uncertainty propagated from the data.}
\label{fig:fit}
\end{figure}

Then the BESIII data from Ref.~\cite{BESIII:2022ner} are fitted using Eq.~\eqref{eq:xsection} with $T_{11}$ from Eq.~\eqref{eq:t11v2} and with the three free parameters in the fit being $\{{\cal N},R,a_{22}\}$---their fitted values are listed in Table~\ref{tab:fits}. The line shapes for the three best fits are depicted in Fig.~\ref{fig:fit}. A well-pronounced dip at the $D\bar D^*$ threshold is clearly seen in all three line shapes. 

\begin{table*}[t!]
\begin{ruledtabular}
\caption{Parameters of the best fits to the BESIII data~\cite{BESIII:2022ner}. The uncertainties are propagated from the experimental data only.
Note that, while $a_{22}$ was fitted to data, 
$a_{11}$ and $|a_{12}|$ were then computed from $a_{22}$ and the value of $a_{22,\text{eff}}$ in Eq.~\eqref{eq:a22effN}.
}\label{tab:fits}
\begin{tabular}{ccccc | c c}
& ${\cal N}$ (pb)& $R\times 10^2$& $a_{22}$ (fm) &$\chi^2/\mbox{dof}$~ & $a_{11}$ (fm) & $|a_{12}|$ (fm)~ \\
\hline
Fit 1 &$2.6_{-1.1}^{+1.4}$&$0.27_{-0.29}^{+0.34}$&$-6.6_{-2.0}^{+2.8}$&0.02
& $-0.51_{-0.22}^{+0.25} $ & $1.8_{-0.8}^{+0.5}$ \\
Fit 2 &$0.18_{-0.07}^{+0.09}$&$5.9_{-2.3}^{+3.2}$&$-10.8_{-6.6}^{+2.0}$&0.18 & $-1.0_{-1.7}^{+0.3} $ & $2.8_{-0.4}^{+0.7}$\\
Fit 3 &$0.41_{-0.16}^{+0.23}$&$-2.6_{-1.5}^{+1.0}$&$-12.8_{-13.2}^{+\phantom{1}3.2}$&0.15 & $-1.4_{-19.8}^{+\phantom{1}0.5} $ & $3.1_{-0.5}^{+0.6}$
\end{tabular}
\end{ruledtabular}
\end{table*}

We conclude that the measured $X(3872)$ line shape is well described by the $J/\psi\rho^0$-$D\bar D^*$ coupled-channel rescattering mechanism outlined above. Although the parameters of the three fits differ substantially from each other, all fits provide equally decent description of the data and possess common gross features. In particular, the three line shapes are nearly indistinguishable 
in the proximity of the $D\bar{D}^*$ threshold 
where they all show
a pronounced dip. The large absolute and negative values of $a_{22}$ found in all three fits imply a 
loosely
bound $D\bar D^*$ state in the single-channel case. Additional data
in the vicinity of the $D\bar{D}^*$ threshold would
allow us to better constrain the model and extract the interaction strengths in different channels with higher precision. 
Also, we emphasise that, while particular details of the line shape in the $e^+e^-\to J/\psi\pi^+\pi^-$ production reaction depend on a delicate interplay of several parameters, the main mechanism driving the dip at the $D\bar{D}^*$ threshold is 
general and is controlled by the large scattering length in this channel.

At leading order in 
heavy quark spin symmetry, the $D\bar D^*$ interaction with $J^{PC}=1^{++}$
agrees with that of 
$D^*\bar D^*$ with $J^{PC}=2^{++}$. Accordingly, if the former interaction generates the $X(3872)$, the latter is predicted 
to generate its
spin-2 partner state~\cite{Nieves:2012tt,Guo:2013sya,Baru:2016iwj}.
We therefore predict an analogous 
dip as discussed above 
near the $D^*\bar D^*$
threshold in the production process $e^+e^-\to J/\psi\pi^+\pi^-\pi^0$.
We notice that the cross section of $e^+e^-\to X_2(4014)$ has been estimated to be $\mathcal{O}(10~{\rm pb})$ under the assumption that the $X_2(4014)$ is a
$D^*\bar D^*$ molecule \cite{Shi:2023ntq}. Given that, the predicted dip may be observed at the upcoming BEPC II-U~\cite{BESIII:2020nme}, which will have a luminosity three times that of BEPC II, and at the Super $\tau$-Charm Facility~\cite{Achasov:2023gey}.

\begin{acknowledgements}
We would like to thank Henryk Czy\.z, Achim Denig, Meng-Lin Du, and Chang-Zheng Yuan for useful discussions. This work is supported in part by the National Key R\&D Program of China under Grant No. 2023YFA1606703; by the Chinese Academy of Sciences (CAS) under Grant No.~YSBR-101;
by the National Natural Science Foundation of China (NSFC) under Grants No. 12125507, No. 12361141819, and No. 12047503; and by NSFC and the Deutsche Forschungsgemeinschaft (DFG) through the funds provided to the Sino-German Collaborative Research Center CRC110 ``Symmetries and the Emergence of Structure in QCD'' (DFG Project No. 196253076).
Work of A.N. was supported by the Slovenian Research Agency (research core Funding No. P1-0035) and the CAS President’s International Fellowship Initiative (PIFI) (Grant No. 2024PVA0004).
\end{acknowledgements}

\bibliographystyle{apsrev4-2}
\bibliography{refs.bib}

\end{document}